\documentclass[12pt]{article}

\usepackage{a4,epsfig,amssymb}
\divide\textwidth\mag \multiply\textwidth by 1000
\divide\textheight\mag \multiply\textheight by 1000
\newcommand{\Real}{\mathop{\rm Re}\nolimits}
\newcommand{\Imag}{\mathop{\rm Im}\nolimits}
\newcommand{\ein}{{\rm e}}
\newcommand{\iim}{{\rm i}}
\makeatletter
\renewcommand{\@makecaption}[2]{%
	 \vskip 10\p@
	 \setbox\@tempboxa\hbox{#1. #2}%
	 \ifdim \wd\@tempboxa >\hsize
			 #1. #2\par
		 \else
			 \hbox to\hsize{\hfil\box\@tempboxa\hfil}%
	 \fi}
\renewcommand\maketitle{\par
 \begingroup
   \def\thefootnote{\fnsymbol{footnote}}%
   \def\@makefnmark{\hbox
       to\z@{$\m@th^{\@thefnmark}$\hss}}%
   \if@twocolumn
     \twocolumn[\@maketitle]%
     \else \newpage
     \global\@topnum\z@
     \@maketitle \fi\@thanks
 \endgroup
 \setcounter{footnote}{0}%
 \let\maketitle\relax
 \let\@maketitle\relax
 \gdef\@thanks{}\gdef\@author{}\gdef\@title{}\let\thanks\relax}
\renewcommand{\section}{\@startsection{section}{1}{0pt}%
{5.5ex plus 1ex minus 0ex}{2.3ex plus .2ex}{\Large\bf}}
\renewcommand{\subsection}{\@startsection{subsection}{2}{0pt}%
{5.5ex plus 1ex minus 0ex}{2.3ex plus .2ex}{\large\bf}}
\renewcommand{\@biblabel}[1]{#1.\ }
\renewcommand{\@oddhead}{\hfil \hfil}
\renewcommand{\@oddfoot}{\hfill \thepage}
\makeatother
\begin{document}

\sloppy
\raggedbottom

\title{\vspace*{-2.5em}\bf Quantum Electron Plasma and Interaction
of S-wave with Thin Metallic Film \vspace*{.5em}}

\author{A.A. Yushkanov$^1$ and N.V. Zverev$^2$ \vspace*{.5em}
\and
\sl Faculty of Physics and Mathematics, \\
\sl Moscow Regional State University, \\
\sl Radio str. 10a, 100500 Moscow, Russia \\
}

\footnotetext[1]{\ \ yushkanov@inbox.ru}
\footnotetext[2]{\ \ zverev\_nv@mail.ru}

\date{}

\maketitle

\vspace*{-2em}

\begin{abstract}
{\normalsize
An interaction of electromagnetic S-wave with thin flat metallic film
is numerically studied for the quantum degenerate electron plasma. One
considers the reflectance, transmittance and absorptance power
coefficients. A contribution of quantum wave properties of electrons
to the power coefficients is shown in comparison of the coefficients
with ones evaluated for both the classical spatial dispersion and the
Drude -- Lorentz approaches. This contribution is detected for the
infrared and terahertz frequencies in case of nanoscale film width.

\vspace*{.5em}


\vspace*{.5em}

}
\end{abstract}

\section*{Introduction}

At current time, a large attention pays to study of the interaction of
electromagnetic waves with tiny or nanoscale metallic films \cite{FuKlPa}
-- \cite{LaYu2}. Such researches have not only theoretical interest
but aimed also on the practical applications in contemporary optical
facilities. Investigations of the interaction widely exploit the kinetic
Fermi -- Dirac electron gas theory. The theory leads to the optical
spatial dispersion property of the electron plasma. And consequently, a
theory of interaction of the electromagnetic wave with the flat infinite
metallic film when the conductivity electrons in metal reflect specularly
from the borders of the film, was elaborated successfully
\cite{KlFu1,JoKlFu}.

But such investigations almost always neglect the de Broglie, or
quantum wave, properties of electrons in the electron plasma. The
principal problem was in getting the correct dielectric functions,
or permittivities, of the quantum electron plasma \cite{Lndh,Mrm},
\cite{LaYu3} -- \cite{LaYu6}. The most pragmatic way to obtain the
functions is when one uses the Liouville -- Schroedinger equation in the
relaxation time approximation for the electron density matrix complying
with conservation laws \cite{Atw}. The second strong difficulty in the
incorporation of the electron wave property into the electron plasma is
an account for the various boundary conditions applied to the electron
density matrix. The well-known specular-diffuse boundary conditions for
the classical kinetic function are not transmitted straightforward to
the density matrix owing to the uncertainty principle. Here, the Wigner
function approaches the classical kinetic one, and common-type classical
boundary conditions due to the Fourier transform are too complicate in
case of quantum plasma (see also \cite{Frnsl}).

However, the electrons in a metal obey the quantum laws. And therefore,
the quantum wave electron effects should influence on light interaction
with a metal. In the paper \cite{YuZv}, the influence of the quantum
wave effects of electron plasma on the interaction of P-wave with
metallic film was shown in case of visible and ultraviolet light.

In this paper, we study the interaction of electromagnetic S-wave with
quantum degenerate electron plasma in the thin flat metallic film placed
between two transparent nonconducting media. We took for investigation
the reflectance, transmittance and absorptance power coefficients. We
consider the quantum degenerate electron plasma with invariable
relaxation time in case of specular electron reflection from the film
surface. The dielectric function (permittivity) of the quantum electron
plasma is taken in the Mermin approach \cite{Mrm,LaYu5}. We investigate
the power coefficients as functions of frequency and of incidence angle.
The evaluated coefficients are compared with those obtained both in case
of the Drude -- Lorentz theory without spatial dispersion and in case of
the classical degenerate electron plasma approach when taking into
account the spatial dispersion.

\section{The model and the power coefficients}

\begin{figure}[ht]

\vspace*{0mm}
\hspace*{10mm}
\epsfig{file=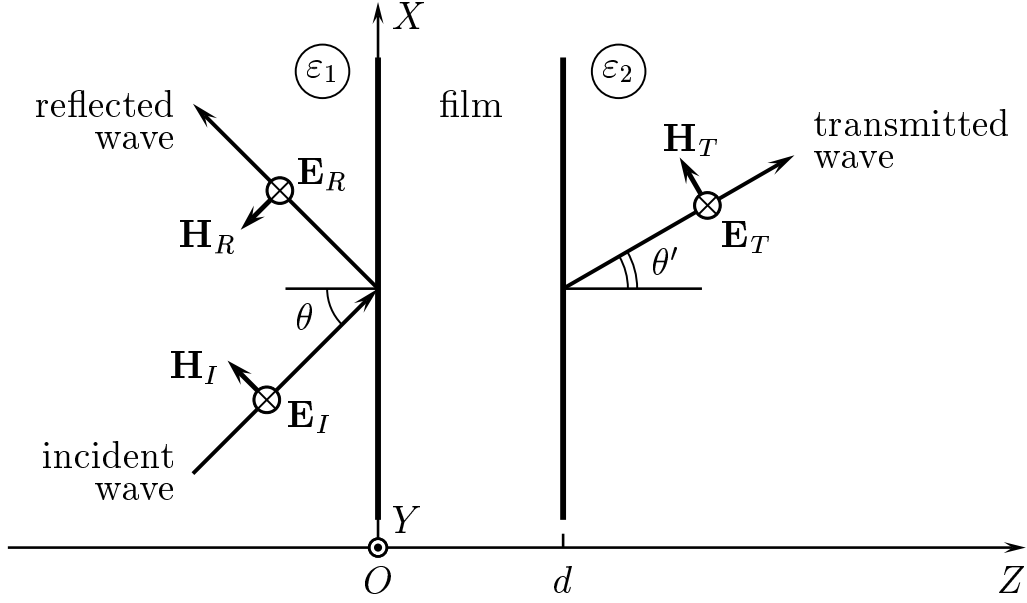,width=.85\textwidth}

\vspace*{0mm}

\caption{The metallic film between two dielectric media with
$\varepsilon_1$ and $\varepsilon_2$, and the incident,
reflected and transmitted waves.}
\label{fig:mtflm-s}

\end{figure}

We consider the flat uniform metallic film of the thickness $d$ placed
between two transparent dielectric media. These media supposed to be
uniform, isotropic and nonmagnetic ones having the positive constant
permittivities $\varepsilon_1$ and $\varepsilon_2$. Hence we neglect
the light dispersion and absorption of the media. Let us suppose that
the electromagnetic wave is incident from the first dielectric medium
on the film under the angle $\theta$ from the surface normal
(fig.~\ref{fig:mtflm-s}). Then in case of a solid second medium, it
can be treated as a substrate.

Let the $Z$ axis is directed orthogonally to the film surface towards
the second dielectric medium. We took $z = 0$ plane as the film surface
contacting with first medium. And hence, the $z = d$ is the second film
surface having contact with the second medium (fig.~\ref{fig:mtflm-s}).
Further, we take the $X$ axis lying both in the film surface and in the
incidence plane towards the wave propagation. And at the end, the
direction of the third $Y$ axis taken in such a way that the rectangular
system is the right-handed one.

We study the S-waves when the ${\bf H}$ vectors of the waves
incident on (${\bf H}_I$), reflected from (${\bf H}_R$) and
transmitted through the film (${\bf H}_T$) lie in the incidence
plane (see fig.~\ref{fig:mtflm-s}). Then the ${\bf E}$ vectors
of the waves (${\bf E}_I$, ${\bf E}_R$ and ${\bf E}_T$) are
parallel to the $Y$ axis.

The electric and magnetic fields of the S-waves in the $z < 0$
half-space projected onto $X$ and $Y$ axes look as \cite{KlFu1,DrGr}
(fig.~\ref{fig:mtflm-s})
\begin{equation}\label{em_1}
\left\lbrace
\begin{array}{lll}
E_y(x,y,z,t) & = & \ein^{\iim(k_{1x}x -\omega t)}\bigl[-a_I\,%
\ein^{\iim k_{1z}z} - a_R\,\ein^{-\iim k_{1z}z}\bigr],
\\
H_x(x,y,z,t) & = & \displaystyle\ein^{\iim(k_{1x}x -\omega t)}%
\bigl[a_I\,\ein^{\iim k_{1z}z} - a_R\,\ein^{-\iim k_{1z}z}\bigr]%
\frac{\sqrt{\varepsilon_1}}{Z_0}\cos\theta.
\end{array}
\right.
\end{equation}

In the $z > d$ half-space, the projections of these fields are
the following:
\begin{equation}\label{em_2}
\left\lbrace
\begin{array}{lll}
E_y(x,y,z,t) & = & -\ein^{\iim(k_{2x}x -\omega t)}%
a_T\,\ein^{\iim k_{2z}(z - d)},
\\
H_x(x,y,z,t) & = & \displaystyle \ein^{\iim(k_{2x}x -\omega t)}a_T%
\frac{\sqrt{\varepsilon_2}}{Z_0}\,\ein^{\iim k_{2z}(z - d)}\cos\theta'.
\end{array}
\right.
\end{equation}

Here $\omega$ is the wave frequency, $k_{1x}$ and $k_{1z}$ are the
$x$- and $z$-projections of the incident wave vector ${\bf k}_1$ in
the initial dielectric medium, $k_{2x}$ and $k_{2z}$ are the same
coordinates of the transmitted wave vector ${\bf k}_2$ in the second
dielectric medium:
\begin{eqnarray}\label{kx}
k_{1x} = \frac{\omega}{c}\sqrt{\varepsilon_1}\sin\theta = k_{2x} =
\frac{\omega}{c}\sqrt{\varepsilon_2}\sin\theta';
\\
k_{1z} = \frac{\omega}{c}\sqrt{\varepsilon_1}\cos\theta, \qquad
k_{2z} = \frac{\omega}{c}\sqrt{\varepsilon_2}\cos\theta'.
\nonumber
\end{eqnarray}
Further, the $a_R$, $a_I$ and $a_T$ stand for the complex electric
field amplitudes of the incident, reflected and transmitted waves,
respectively. The $c$ is the vacuum speed of light, $Z_0$ denotes the
dimensional (in Ohm) vacuum impedance. And at the end, $\theta'$ is
the narrow refraction angle into the second dielectric medium from
the surface normal (fig.~\ref{fig:mtflm-s}). Note that in the case of
total internal reflection when \ $\sin\theta' > 1$, \ the $\cos\theta'$
value is pure imaginary. Then since the transmitted wave should not
amplify infinitely, the sign of the imaginary value has to be
positive i.e. \ $\Imag\cos\theta' > 0$. Therefore, the $\cos\theta'$
is evaluated according to the equation following from (\ref{kx}) one:
\begin{equation}\label{csthpr}
\cos\theta' = \left\lbrace
\begin{array}{lll}
\displaystyle \sqrt{1 - \frac{\varepsilon_1}{\varepsilon_2}%
\sin^2\theta}, & & \displaystyle \sin\theta \leqslant
\sqrt{\frac{\varepsilon_2}{\varepsilon_1}};
\\[1em]
\displaystyle \iim\sqrt{\frac{\varepsilon_1}{\varepsilon_2}%
\sin^2\theta - 1}, & & \displaystyle \sin\theta >
\sqrt{\frac{\varepsilon_2}{\varepsilon_1}}.
\end{array}
\right.
\end{equation}

In the film zone $0 < z < d$, the electric and magnetic fields may be
represented in the following way:
\begin{equation}\label{em_sl}
\left\lbrace
\begin{array}{lll}
E_y(x,y,z,t) & = & \ein^{\iim(k_{1x}x -\omega t)}\bigl[\alpha_1
E^{(1)}_y(z) + \alpha_2 E^{(2)}_y(z)\bigr],
\\
H_x(x,y,z,t) & = & \ein^{\iim(k_{1x}x -\omega t)}\bigl[\alpha_1
H^{(1)}_x(z) + \alpha_2 H^{(2)}_x(z)\bigr].
\end{array}
\right.
\end{equation}
Here $\alpha_1$ and $\alpha_2$ are some constant coefficients, and
$E^{(j)}_y(z)$ and $H^{(j)}_x(z)$ stand for the antisymmetric or
symmetric modes of electric and magnetic fields ($j = 1,2$):
\begin{equation}\label{em_md}
E^{(j)}_y(z) = (-1)^j E^{(j)}_y(d - z), \qquad
H^{(j)}_x(z) = (-1)^{j+1} H^{(j)}_x(d - z).
\end{equation}

The electric and magnetic fields on the film surfaces $z = 0$ and
$z = d$ have to satisfy the boundary conditions:
\begin{equation}\label{em_bc1}
\left\lbrace
\begin{array}{lll}
E_y(x,y,-0,t) & = & E_y(x,y,+0,t),
\\
H_x(x,y,-0,t) & = & H_x(x,y,+0,t);
\end{array}
\right.
\end{equation}
\begin{equation}\label{em_bc2}
\left\lbrace
\begin{array}{lll}
E_y(x,y,d-0,t) & = & E_y(x,y,d+0,t),
\\
H_x(x,y,d-0,t) & = & H_x(x,y,d+0,t).
\end{array}
\right.
\end{equation}

In order to connect the electric and magnetic fields on the film
surfaces, it is convenient to use the dimensionless surface impedance
\cite{KlFu1} -- \cite{KlFu3}. For the S-wave modes on the $z = 0$
surface, the impedance defined as ($j = 1,2$):
\begin{equation}\label{sf_imp1}
Z^{(j)}_S = \frac{1}{Z_0}\frac{E^{(j)}_y(+0)}{H^{(j)}_x(+0)}.
\end{equation}

Substituting (\ref{em_1}) and (\ref{em_sl}) into (\ref{em_bc1}),
(\ref{em_2}) and (\ref{em_sl}) into (\ref{em_bc2}), employing the
property (\ref{em_md}) of the modes and using the surface impedance
(\ref{sf_imp1}), after elimination of similar multipliers one gets
the following system:
\begin{equation}\label{em_sys1}
\left\lbrace
\begin{array}{rcl}
-(a_I + a_R) & = & \alpha_1 E^{(1)}_y(+0) + \alpha_2 E^{(2)}_y(+0),
\\
(a_I - a_R)\sqrt{\varepsilon_1}\cos\theta & = & \displaystyle
\alpha_1\frac{E^{(1)}_y(+0)}{Z^{(1)}_S} +
\alpha_2\frac{E^{(2)}_y(+0)}{Z^{(2)}_S},
\\[1em]
-a_T & = & -\alpha_1 E^{(1)}_y(+0) + \alpha_2 E^{(2)}_y(+0),
\\
a_T\sqrt{\varepsilon_2}\cos\theta' & = & \displaystyle
\alpha_1\frac{E^{(1)}_y(+0)}{Z^{(1)}_S} -
\alpha_2\frac{E^{(2)}_y(+0)}{Z^{(2)}_S}.
\end{array}
\right.
\end{equation}

After elimination of the values $\alpha_j E^{(j)}_y(+0)$ ($j = 1,2$)
from the system (\ref{em_sys1}), one comes to the system
\begin{equation}\label{em_sys2}
\left\lbrace
\begin{array}{rcl}
a_T & = & U^{(1)}_S a_I + V^{(1)}_S a_R,
\\
a_T & = & -U^{(2)}_S a_I - V^{(2)}_S a_R.
\end{array}
\right.
\end{equation}
Here is denoted ($j = 1,2$):
\begin{equation}\label{uv_c}
U^{(j)}_S = \frac{1 + Z^{(j)}_S\sqrt{\varepsilon_1}\cos\theta}%
{1 - Z^{(j)}_S\sqrt{\varepsilon_2}\cos\theta'}, \qquad
V^{(j)}_S = \frac{1 - Z^{(j)}_S\sqrt{\varepsilon_1}\cos\theta}%
{1 - Z^{(j)}_S\sqrt{\varepsilon_2}\cos\theta'}.
\end{equation}

\bigskip

Let us turn now to the definition of the reflectance $R$, transmittance
$T$ and absorptance $A$ power coefficients. The first two of them are
the ratios \cite{LdLf,DrGr}:
\begin{equation}\label{rt_def}
R = \frac{|\langle S_z\rangle_R|}{|\langle S_z\rangle_I|}, \qquad
T = \frac{|\langle S_z\rangle_T|}{|\langle S_z\rangle_I|},
\end{equation}
where $\langle S_z\rangle$ is the time averaged energy flux density,
or Poynting, vector, projected onto the $Z$ axis:
\begin{equation}\label{Pnt1}
\langle S_z\rangle = \frac{1}{2}\Real({\bf E}\times{\bf H}^*)\cdot
{\bf e}_z,
\end{equation}
where $^*$ is the complex conjugation and ${\bf e}_z$ stands for the
unit vector towards the $Z$ axis direction. The $I$, $R$ and $T$
subscript letters in the equations (\ref{rt_def}) stand for the
respective incident, reflected and transmitted waves. The expression
(\ref{Pnt1}) for the S-wave can be transformed to the following one:
\begin{equation}\label{Pnt2}
\langle S_z\rangle = -\frac{1}{2}\Real(E_y H^*_x).
\end{equation}

We substitute the equations (\ref{em_1}) and (\ref{em_2}) to the
(\ref{Pnt2}) and separate the terms related to $\langle S_z\rangle_I$,
$\langle S_z\rangle_R$ and $\langle S_z\rangle_T$. Then we substitute
these terms to the equations (\ref{rt_def}) and use the positivity of
the dielectric constants $\varepsilon_1$ and $\varepsilon_2$. And one
comes to the following expressions for the reflectance $R$ and
transmittance $T$ power coefficients:
\begin{equation}\label{rt_cf1}
R = \left|\frac{a_R}{a_I}\right|^2, \qquad T = \Real\left(\frac{\cos%
\theta'}{\cos\theta}\sqrt{\frac{\varepsilon_2}{\varepsilon_1}}\,\right)%
\left|\frac{a_T}{a_I}\right|^2.
\end{equation}

Evaluating the ratios $a_R/a_I$ and $a_T/a_I$ from the system
(\ref{em_sys2}) and substituting them into the equations (\ref{rt_cf1}),
one arrives at the final equations for the $R$, $T$ and also for the
absorptance $A$:
\begin{eqnarray}
& & R = \left|\frac{U^{(1)}_S + U^{(2)}_S}{V^{(1)}_S + V^{(2)}_S}%
\right|^2, \label{r_cf}
\\
& & T = \Real\left(\frac{\cos\theta'}{\cos\theta}\sqrt{\frac%
{\varepsilon_2}{\varepsilon_1}}\,\right)\left|\frac{U^{(1)}_S %
V^{(2)}_S - U^{(2)}_S V^{(1)}_S}{V^{(1)}_S + V^{(2)}_S}\right|^2,
\label{t_cf}
\\
& & A = 1 - R - T. \label{a_cf}
\end{eqnarray}

The power coefficients (\ref{r_cf}) -- (\ref{a_cf}) comply with those
obtained in paper \cite{LaYu2s} for the S-wave. In the particular case
when the dielectric media are vacuum or air with $\varepsilon_1 = %
\varepsilon_2 = 1$, one has $\theta' = \theta$ and the equations
(\ref{r_cf}), (\ref{t_cf}) reproduce the reflectance and transmittance
power coefficients presented in papers \cite{JoKlFu,KlFu2,KlFu3}.

\section{The surface impedance and the dielectric function of the
degenerate electron plasma}

The surface impedance defined by equation (\ref{sf_imp1}), was evaluated
in \cite{JoKlFu} for the flat uniform metallic film with isotropic
electron plasma in the case of specular electron reflections from the
film borders. It can be rewritten in terms of dimensionless values and
for the S-wave, it looks as follows ($j = 1,2$):
\begin{equation}\label{sf_imp2}
Z^{(j)}_S = -\frac{2\iim\Omega}{\beta W}\sum_n\frac{1}{\Omega^2%
\varepsilon_{tr}(\Omega,Q_n) - (Q_n/\beta)^2},
\end{equation}
where $\Omega$, $\beta$, $W$, $Q_n$, $Q_x$ are the dimensionless
variables and parameters:
\begin{eqnarray}
& & \Omega = \frac{\omega}{\omega_p}, \qquad \beta = \frac{v_F}{c},
\qquad W = \frac{\omega_p\,d}{v_F}, \label{ombtw}
\\
& & Q_n = \sqrt{\Bigl(\frac{\pi n}{W}\Bigr)^2 + Q^2_x}, \qquad
Q_x = \frac{v_F k_x}{\omega_p}. \label{qqx}
\end{eqnarray}
Here $\omega_p$ is the frequency of the degenerate electron plasma,
$v_F$ is the electron Fermi velocity of conductivity electrons, the
$k_x$ is the $x$-projection of the wave vector ${\bf k}$. Further, the
$\varepsilon_{tr}(\Omega,Q)$ is the transverse dielectric function
(permittivity) of the isotropic electron plasma. Summation in
(\ref{sf_imp2}) is performed over all odd integers $n$ if $j = 1$ or
over all even integers if $j = 2$:
$$
\begin{array}{lll}
j = 1: & \ \ & n = \pm 1,\,\pm 3,\,\pm 5,\,\pm 7,\,\ldots;
\\
j = 2: & & n = 0,\,\pm 2,\,\pm 4,\,\pm 6,\,\ldots.
\end{array}
$$

It is convenient to use dimensionless variables in the units of the
degenerate electron plasma. The transverse dielectric function of the
quantum electron plasma at zero temperature with invariable relaxation
time owing to electron collisions, obtained in the Mermin approach
involving the electron density matrix in the momentum space, looks as
follows \cite{Mrm,LaYu5}:
\begin{equation}\label{epqu_tr}
\varepsilon^{(qu)}_{tr}(\Omega,Q) = 1 - \frac{1}{\Omega^2}\left(%
1 + \frac{\Omega G(\Omega + \iim\gamma, Q) + \iim\gamma G(0,Q)}%
{\Omega + \iim\gamma}\right).
\end{equation}
Here the function $G$ is defined by the equation:
\begin{eqnarray}
G(\Omega + \iim\gamma,Q) & = & \frac{3}{16r}\Bigl[ B_2(\Omega_+ +
\iim\gamma, Q) - B_2(\Omega_- + \iim\gamma, Q)\Bigr] +
\nonumber\\
 & + & \frac{9}{8}\Bigl(\frac{\Omega + \iim\gamma}{Q}\Bigr)^2 +
\frac{3}{32}(Qr)^2 - \frac{5}{8}, \nonumber
\end{eqnarray}
where the functions and variables ($\alpha = 1,2$)
\begin{eqnarray}
& & B_{\alpha}(\Omega + \iim\gamma, Q) = \frac{1}{Q^{2\alpha+1}}\bigl[%
(\Omega + \iim\gamma)^2 - Q^2\bigr]^{\alpha}\,L(\Omega + \iim\gamma, Q),
\label{b12} \\
& & L(\Omega + \iim\gamma, Q) = \ln\frac{\Omega + \iim\gamma - Q}%
{\Omega + \iim\gamma + Q}, \label{ln_rt}
\\
& & \Omega_{\pm} = \Omega \pm \frac{1}{2}Q^2 r. \nonumber
\end{eqnarray}
And further, the dimensionless variable and parameters
\begin{equation}\label{qgar}
Q = \frac{v_F |{\bf k}|}{\omega_p}, \qquad \gamma =
\frac{1}{\omega_p\tau}, \qquad r = \frac{\hbar\omega_p}{m_e v^2_F}.
\end{equation}
Here $\tau$ is the relaxation time owing to the electron collisions,
$m_e$ is the effective mass of the conductance electrons and $\hbar$ is
the Planck constant.

The complex logarithm ratio $L(\Omega + \iim\gamma, Q)$ defined by
equation (\ref{ln_rt}), has the branching real negative half-line on
the complex plane $\Omega + \iim\gamma$. Then one must evaluate the
logarithm ratio using the prescription:
\begin{equation}\label{lr_evl}
L(\Omega + \iim\gamma, Q) = \frac{1}{2}\ln\frac{(\Omega - Q)^2 + %
\gamma^2}{(\Omega + Q)^2 + \gamma^2} + \iim\left(\arctan\frac{\Omega %
+ Q}{\gamma} - \arctan\frac{\Omega - Q}{\gamma}\right).
\end{equation}
From (\ref{lr_evl}) follows that the function $G(0,Q)$ is real since
$Q\geqslant 0$ and $\gamma\geqslant 0$.

The transverse dielectric function (\ref{epqu_tr}) in the classical limit
$r\to 0$ go over to corresponding dielectric function of the degenerate
Fermi gas disregarding for the quantum wave electron properties called as
classical spatial dispersion case \cite{KlFu1,LaYu2,LaYu2s}:
\begin{eqnarray}
\varepsilon^{(cl)}_{tr}(\Omega,Q) & = & 1 - \frac{3}{4\Omega}\Biggl(%
\frac{2(\Omega + \iim\gamma)}{Q^2} + B_1(\Omega + \iim\gamma, Q)\Biggr),
\label{epcl_tr}
\end{eqnarray}
where the function $B_1(\Omega + \iim\gamma, Q)$ is defined by the
equation (\ref{b12}). And also in the infinite wavelength limit $Q\to 0$,
both the quantum dielectric function (\ref{epqu_tr}) and the classical
spatial dispersion one (\ref{epcl_tr}) go over to the classical Drude
-- Lorentz electron dielectric function without spatial dispersion
\cite{DrGr}:
\begin{equation}\label{epDL}
\varepsilon^{(DL)}_{tr}(\Omega) = 1 - \frac{1}{\Omega(\Omega + %
\iim\gamma)}.
\end{equation}
It worth noting that in the case of Drude -- Lorentz dielectric function
(\ref{epDL}), one can perform exactly the summation in (\ref{sf_imp2}).

We return to the surface impedance (\ref{sf_imp2}). It is used in
(\ref{uv_c}) to evaluate the power coefficients (\ref{r_cf}) --
(\ref{a_cf}). The $Q_x$ variable in $Q_n$ is evaluated by substituting
$k_x = k_{1x}$ by (\ref{kx}) in the (\ref{qqx}) for $Q_x$ and using
the definitions (\ref{ombtw}). Then one obtains:
\begin{equation}\label{qx2}
Q_x = \Omega\beta\sqrt{\varepsilon_1}\sin\theta.
\end{equation}

\section{Numerical studies of the power coefficients
for quantum electron plasma}

We perform numerical investigation of the reflectance $R$, transmittance
$T$ and absorptance $A$ power coefficients for S-wave in case of the
quantum plasma. The coefficients are evaluated according to the equations
(\ref{r_cf}) -- (\ref{a_cf}) when one uses the formulas (\ref{csthpr}),
(\ref{uv_c}), (\ref{sf_imp2}), (\ref{qqx}) and (\ref{qx2}) with the
dielectric function (\ref{epqu_tr}). We study the coefficients as
functions of variable $\Omega$ called by us as dimensionless frequency,
and of the incidence angle $\theta$, for various film widths $d$ and for
different surrounding dielectric media.

The initial data taken by us are characteristics of potassium
\cite{KlFu2}: $\omega_p=6.61\cdot{}10^{15}$~sec$^{-1}$,
\ $v_F=8.5\cdot{}10^5$~m/sec, the effective mass of conductance
electron $m_e$ equals to the free electron mass, and the dimensionless
parameter $\gamma=10^{-3}$. So taking into account (\ref{ombtw})
and (\ref{qgar}), we have set the values $\beta=2.83\cdot{}10^{-3}$ and
$r=1.07$.

\begin{figure}[ht]

\vspace*{27mm}
\hspace*{-5mm}
\epsfig{file=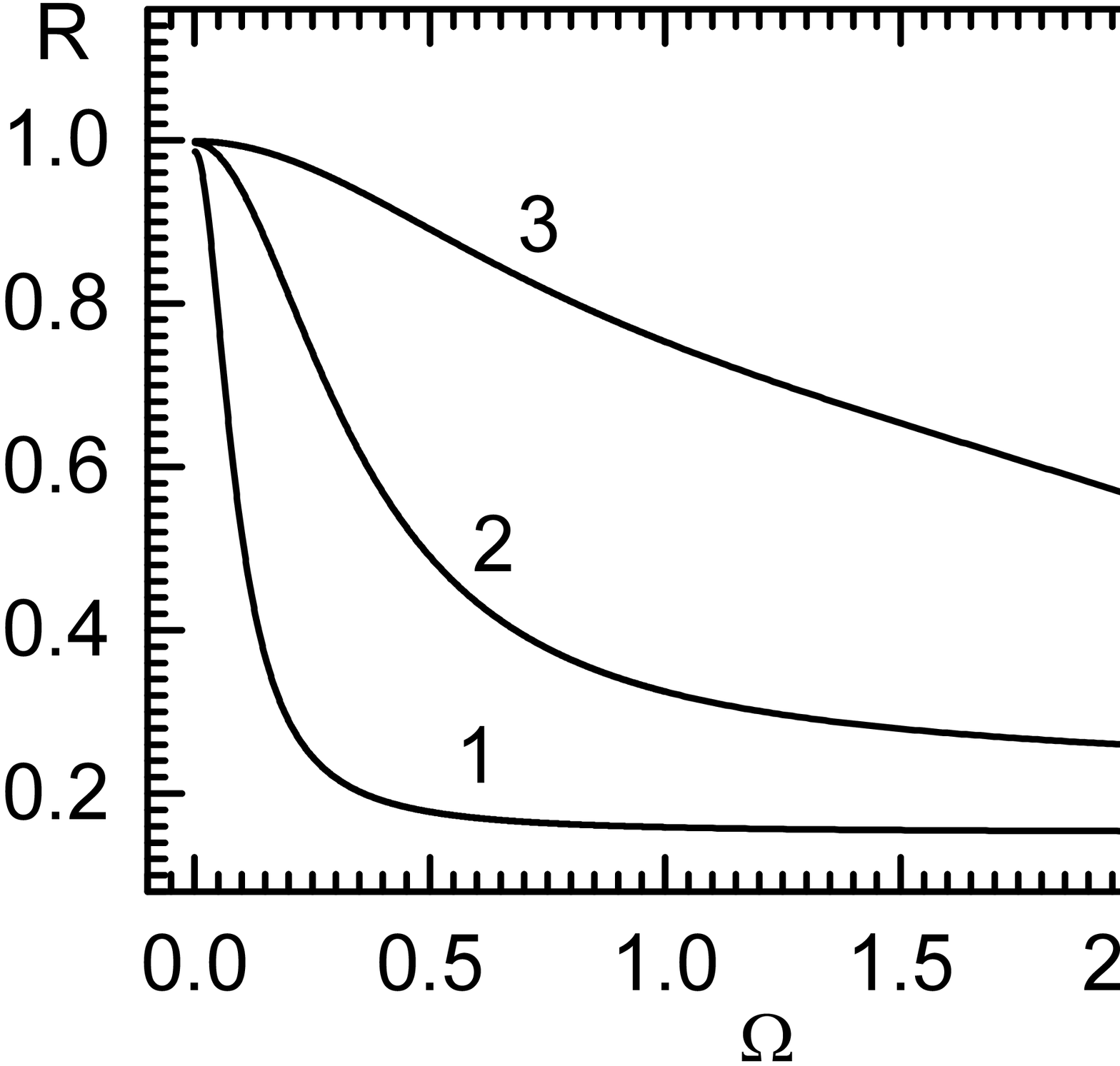,width=.52\textwidth,angle=0}
\hspace*{-5mm}
\epsfig{file=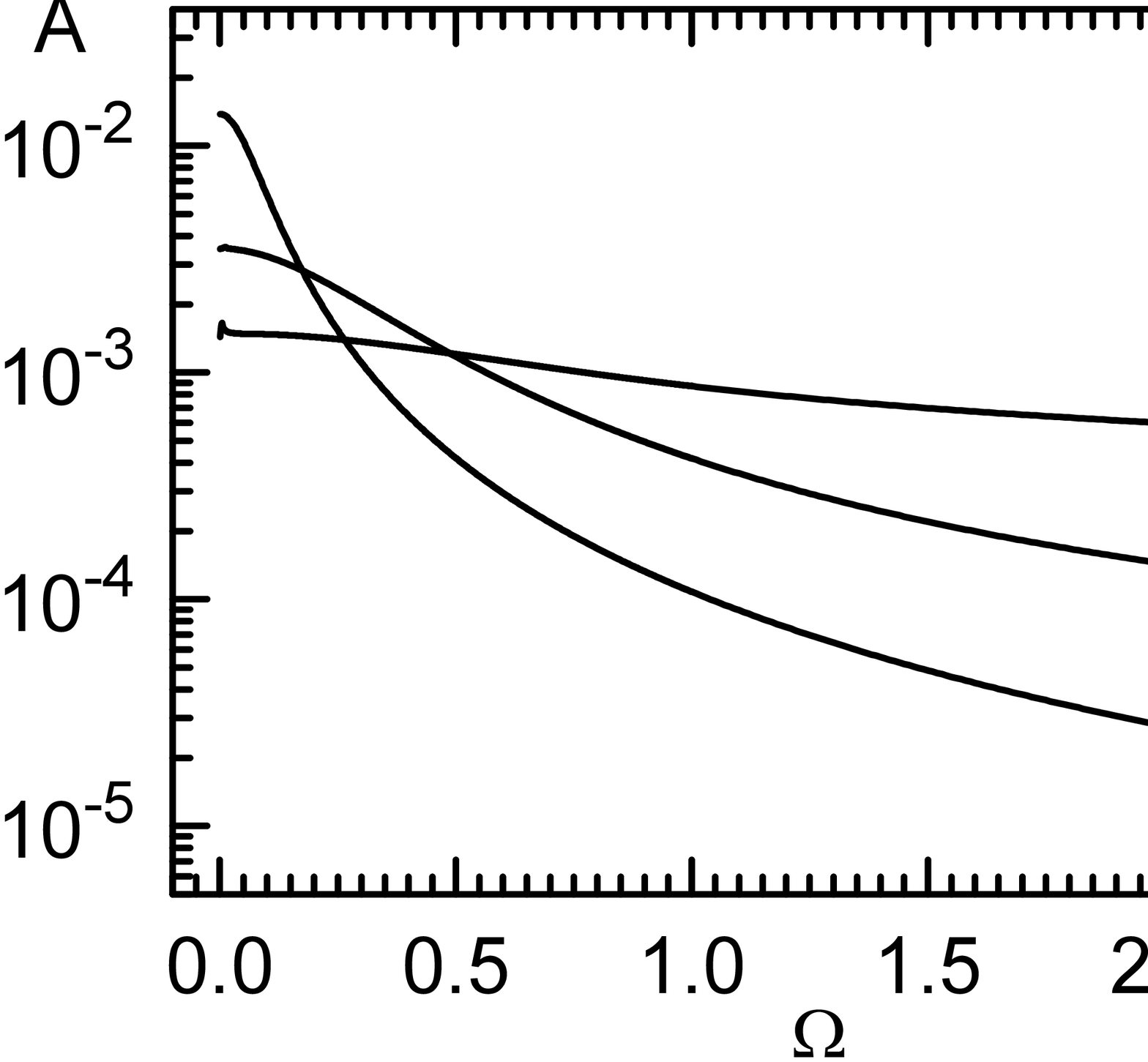,width=.52\textwidth,angle=0}

\vspace*{-27mm}

\caption{The reflectance $R$ (left plot) and absorptance $A$ (right
plot) as functions of $\Omega$ for quantum plasma at $\beta=2.83%
\cdot{}10^{-3}$, $\gamma=10^{-3}$, $r=1.07$, $\theta=60^{\circ}$,
$\varepsilon_1=1$ (air), $\varepsilon_2=2$ (quartz): \ 1 -- $W = 50$,
\ 2 -- $W = 200$, \ 3 -- $W = 500$.}
\label{fig:raq_s}

\end{figure}

Typical results for $R$ and $A$ as functions of dimensionless frequency
$\Omega$ for some values $W$ are shown at the fig.~\ref{fig:raq_s}.
The chosen values $W=50$, $200$ and $500$ correspond according to
(\ref{ombtw}), to the film widths $d=6.43$~nm, $25.72$~nm and $64.3$~nm,
respectively. The first surrounding medium is an air or vacuum with
$\varepsilon_1 = 1$, and the second medium or substrate, is a quartz
with $\varepsilon_2 = 2$. The considered frequency interval
$10^{-3}\leqslant\Omega\leqslant{}2.5$ covers the ranges from the
terahertz range to the ultraviolet one. One sees that almost always
in the selected frequency interval, the reflectance $R$ and absorptance
$A$ power coefficients decrease with growth of frequency. And also,
a decrease of the power coefficients becomes less sharp with an
increase of the film width $d$.

\begin{figure}[ht]

\vspace*{27mm}
\hspace*{-5mm}
\epsfig{file=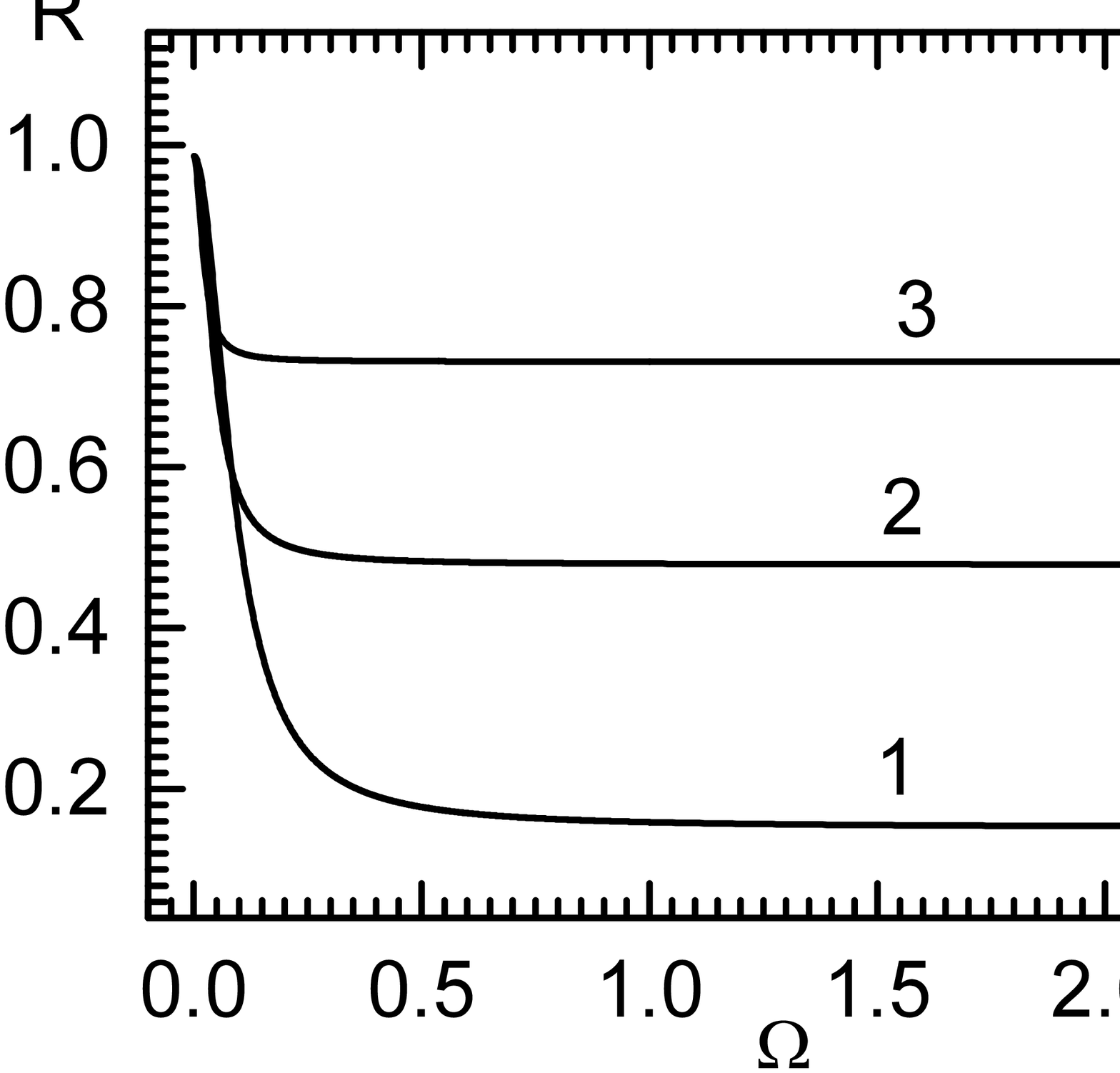,width=.52\textwidth,angle=0}
\hspace*{-5mm}
\epsfig{file=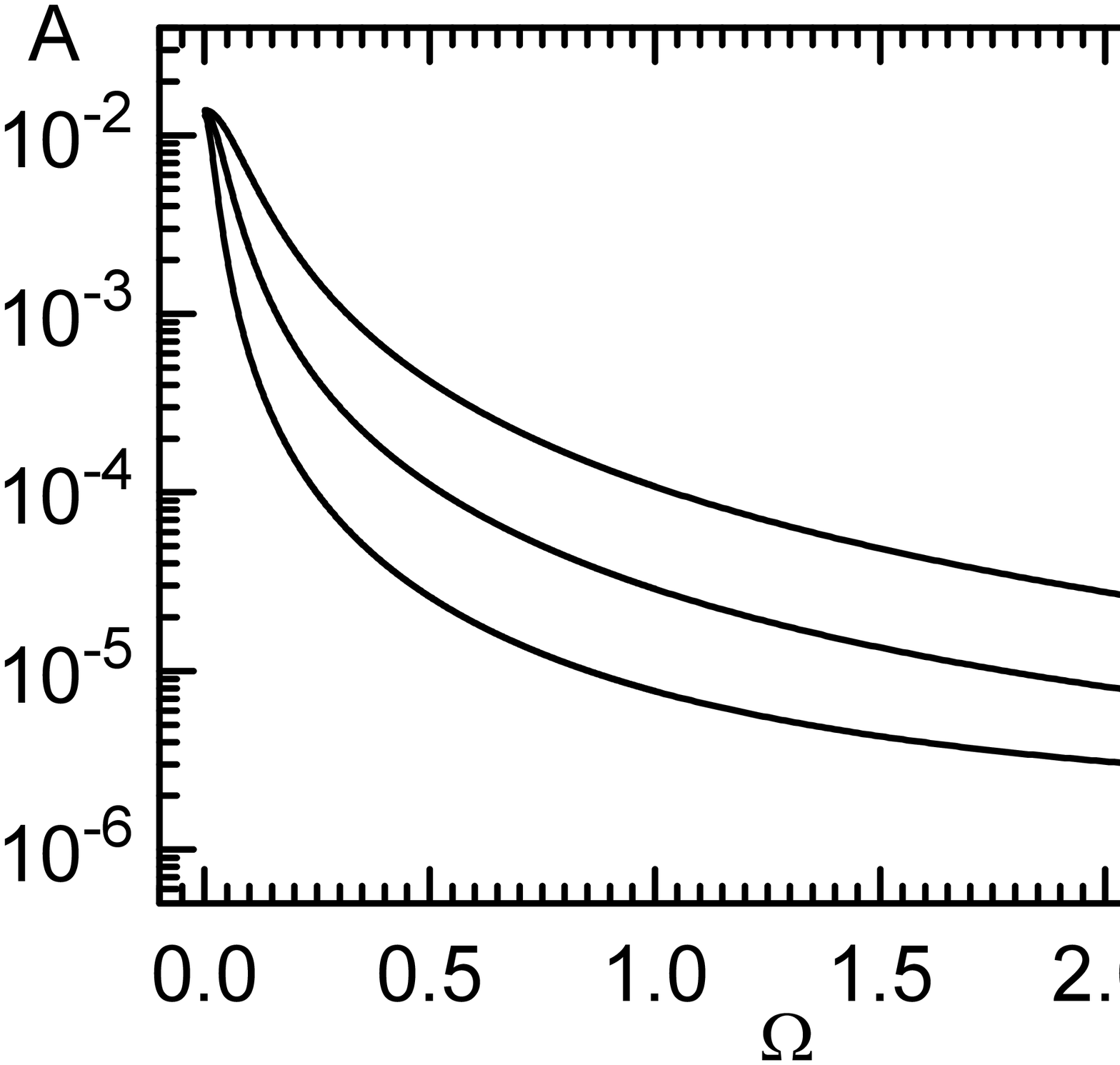,width=.52\textwidth,angle=0}

\vspace*{-27mm}

\caption{The reflectance $R$ (left plot) and absorptance $A$ (right
plot) as functions of $\Omega$ for quantum plasma at $\beta=2.83%
\cdot{}10^{-3}$, $\gamma=10^{-3}$, $r=1.07$, $\theta=60^{\circ}$,
$W=50$, $\varepsilon_1=1$ (air): \ \ 1 -- $\varepsilon_2=2$ (quartz),
\ 2 -- $\varepsilon_2=8$ (mica), \ 3 -- $\varepsilon_2=40$ (technical
ceramics).}
\label{fig:ra284s}

\end{figure}

Further, in the fig.~\ref{fig:ra284s} we show the coefficients $R$ and
$A$ as functions of $\Omega$ for various second media, or substrates:
a quartz with $\varepsilon_2=2$, a mica with $\varepsilon_2=8$, and a
technical ceramics with $\varepsilon_2=40$. It is seen that within the
considered frequency interval, both the reflectance $R$ and absorptance
$A$ coefficients decrease with $\Omega$ growth. And the reflectance
coefficient at the values $\Omega\gtrsim{}0.1$\,---\,$0.25$ almost
goes to a saturation. But if the absorptance decreases more strong
at increase of the dielectric constant $\varepsilon_2$ of the second
medium, the reflectance vice versa, decrease more gentle sloping.

\begin{figure}[ht]

\vspace*{27mm}
\hspace*{-2mm}
\epsfig{file=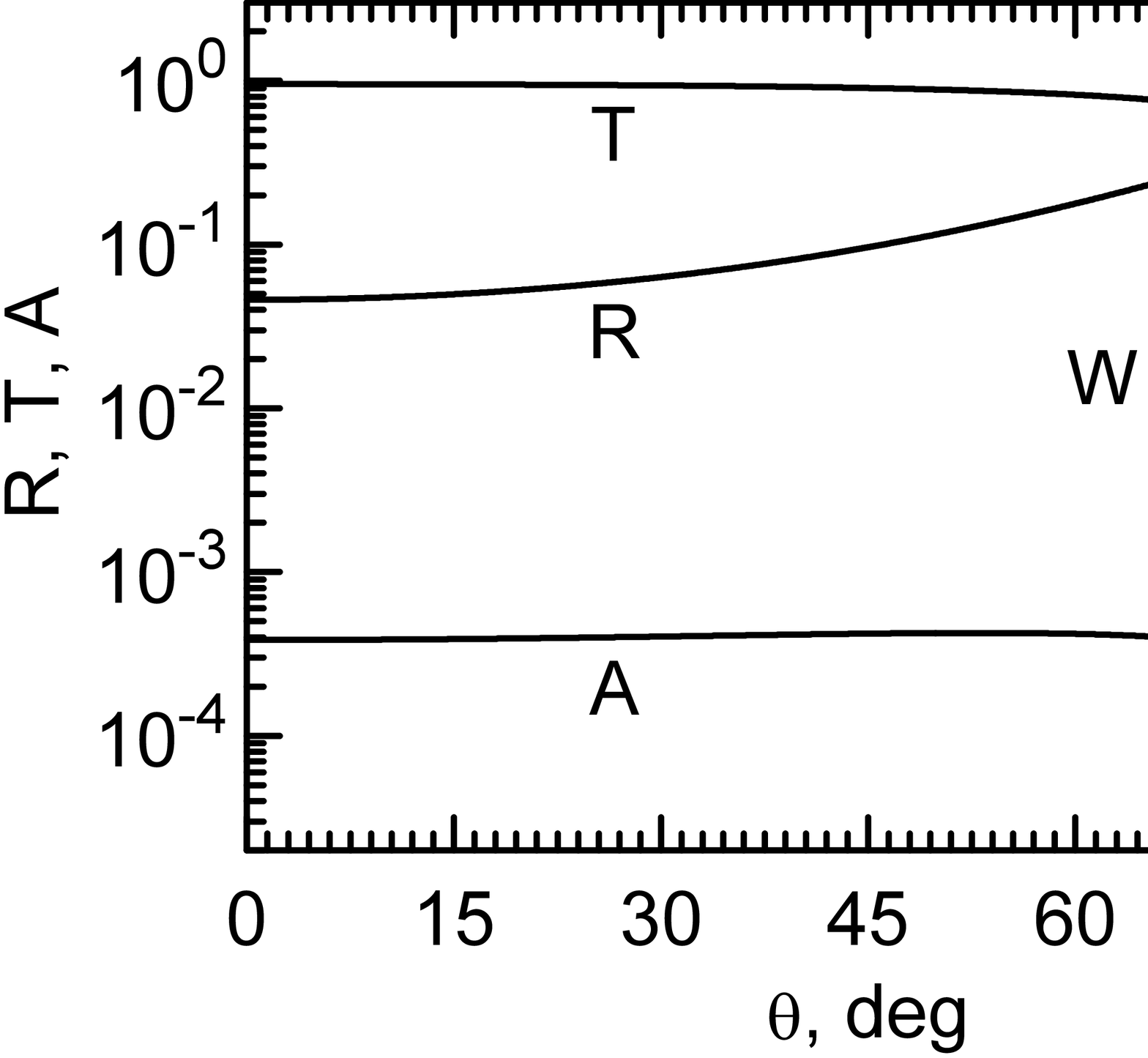,width=.5\textwidth,angle=0}
\hspace*{-3mm}
\epsfig{file=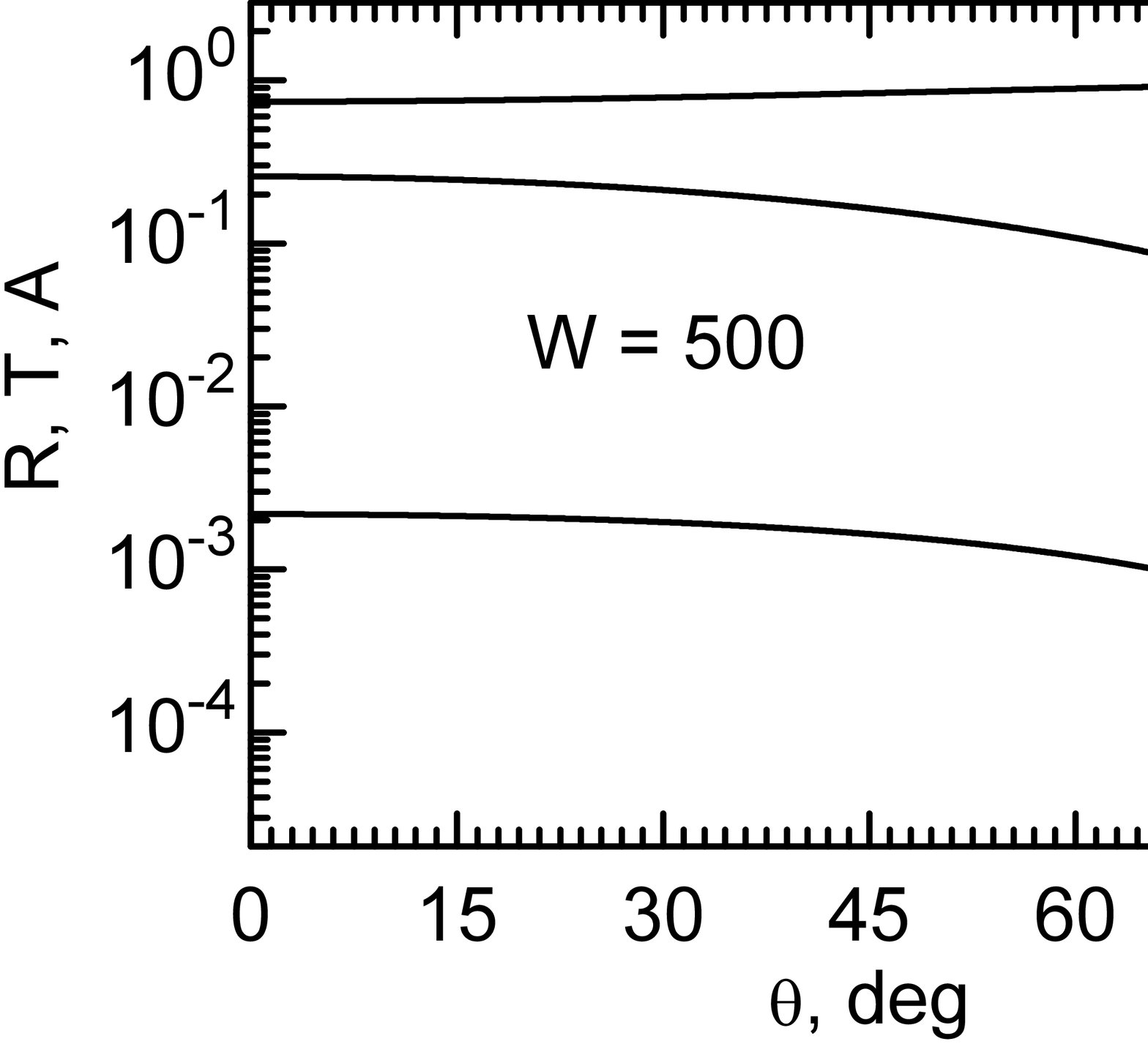,width=.5\textwidth,angle=0}

\vspace*{-27mm}

\caption{The coefficients $R$, $T$ and $A$ as functions of $\theta$
for quantum plasma at $\beta=2.83\cdot{}10^{-3}$, $\gamma=10^{-3}$,
$r=1.07$, $\Omega=0.5$, $\varepsilon_1=1$, $\varepsilon_2=2$, $W=50$
(left plot), $W=500$ (right plot).}
\label{fig:cfthq}

\end{figure}

And at the end of the section, we studied the power coefficients as
functions of the incidence angle $\theta$. Typical results are presented
at the fig.~\ref{fig:cfthq} for two film widths with $W=50$ and $W=500$.
One sees that the reflectance $R$ almost always increases whereas the
transmittance $T$ and absorptance $A$ decrease with growth of $\theta$.

It worth to compare the results for S-wave with those in case of P-wave
when the ${\bf E}$ vectors lie in the incidence plane, obtained in the
quantum plasma approach \cite{YuZv}. The behavior of the power
coefficients as functions of $\theta$ in case of S-wave is in somewhat
reminiscent to the behavior for the P-wave. However in case of the
P-wave, one observes resonant peaks of power coefficients in the region
$\Omega\gtrsim{}1$. These peaks are caused by influence of longitudinal
plasmons moving between borders of the metallic film \cite{JoKlFu,PSChE}.
But here for the S-wave, these oscillations do not occur because the
${\bf E}$ vectors are parallel to film borders. And hence, one observes
a smooth behavior of the power coefficients.

\section{Comparison with classical spatial dispersion and Drude --
Lorentz approaches}

Now let us compare the results for S-wave in case of quantum plasma with
those obtained for the classical spatial dispersion plasma and in case
of Drude -- Lorentz approach. To evaluate the power coefficients in the
latter approaches, one employs again the equations (\ref{r_cf}) --
(\ref{a_cf}) with (\ref{csthpr}), (\ref{uv_c}), (\ref{sf_imp2}),
(\ref{qqx}) and (\ref{qx2}). But now the dielectric function
(\ref{epqu_tr}) of the quantum plasma is replaced either by
(\ref{epcl_tr}) or by (\ref{epDL}) one.

\begin{figure}[ht]

\vspace*{27mm}
\hspace*{-2mm}
\epsfig{file=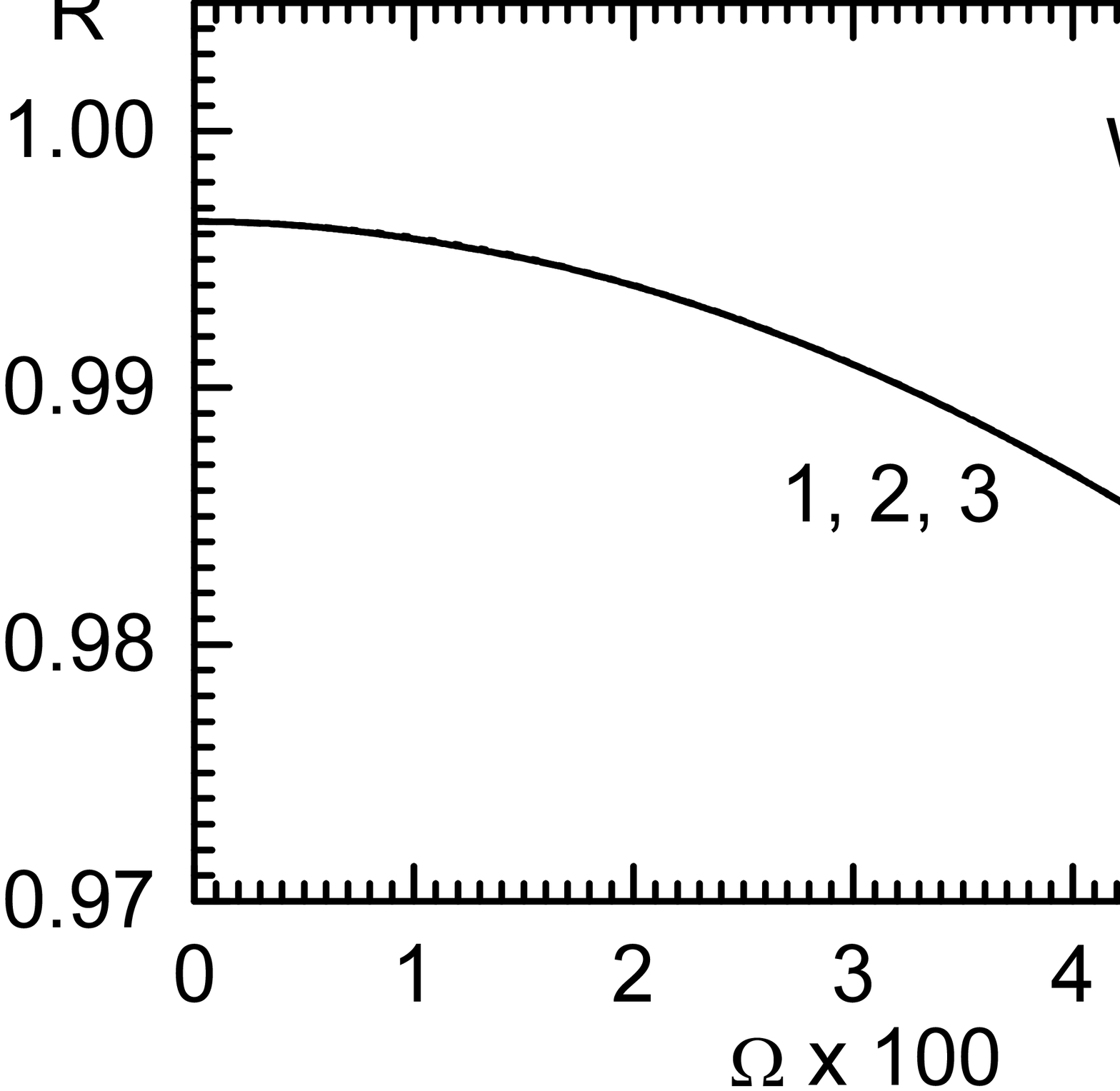,width=.5\textwidth,angle=0}
\hspace*{-1mm}
\epsfig{file=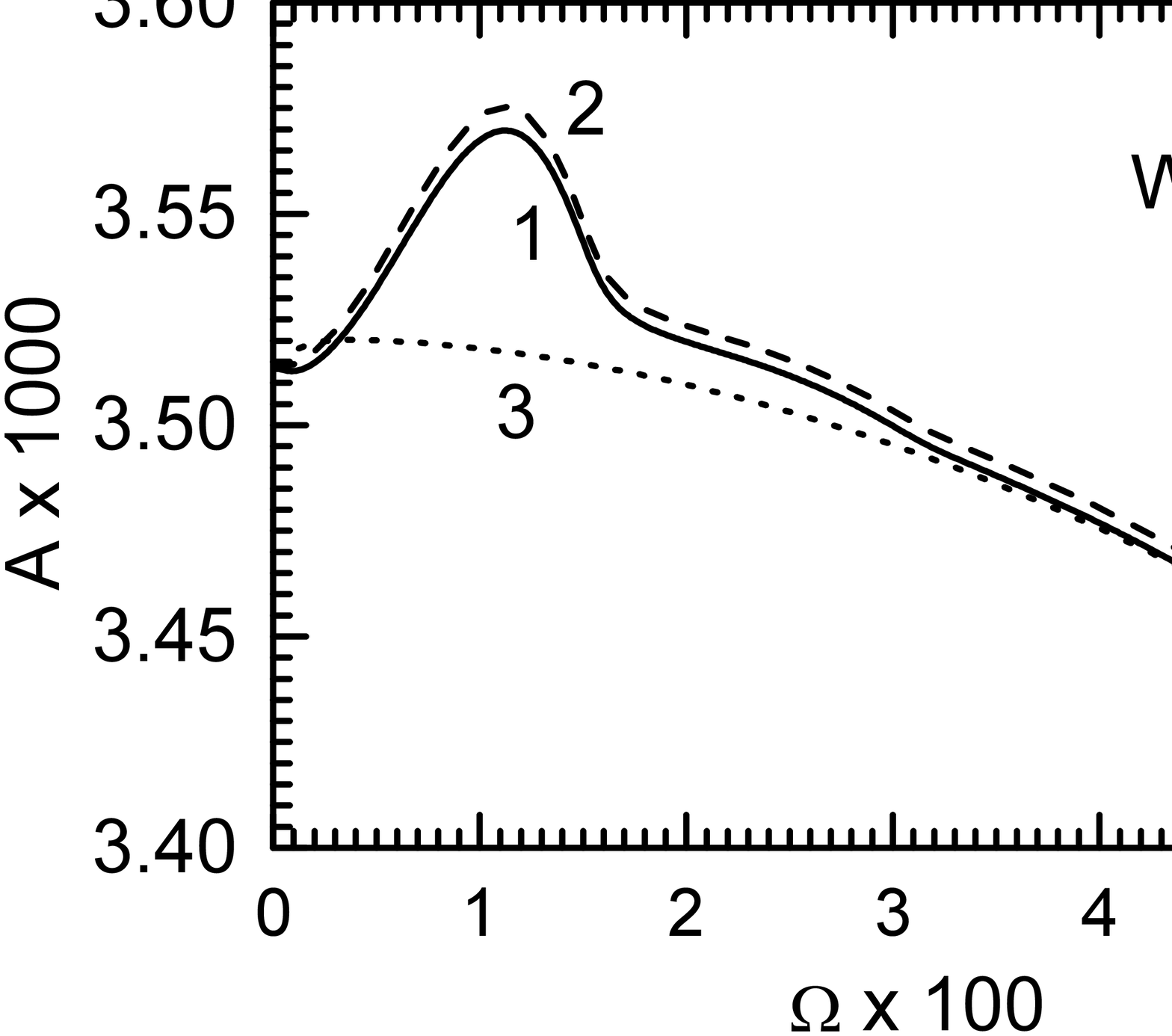,width=.5\textwidth,angle=0}

\vspace*{2mm}
\hspace*{-2mm}
\epsfig{file=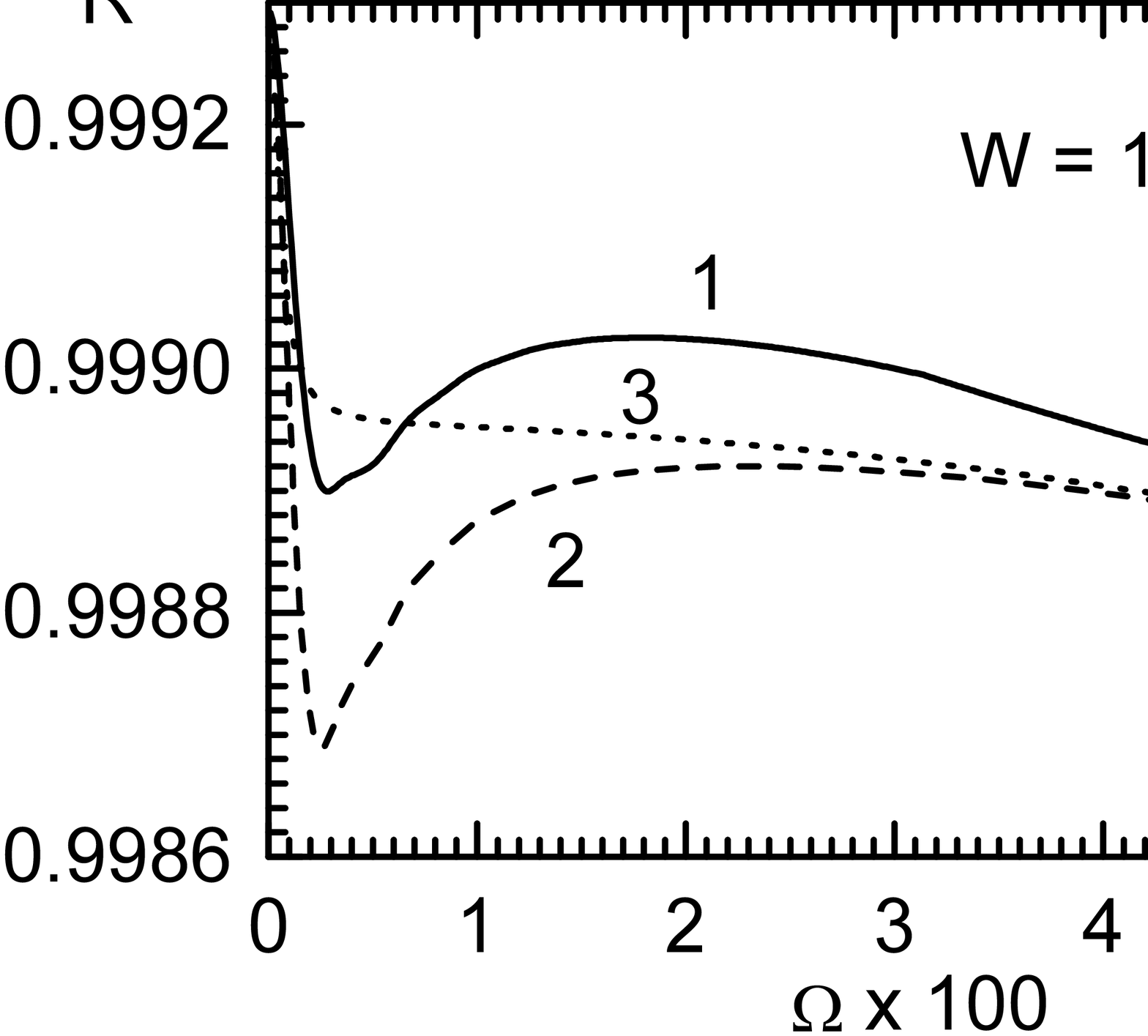,width=.5\textwidth,angle=0}
\hspace*{-1mm}
\epsfig{file=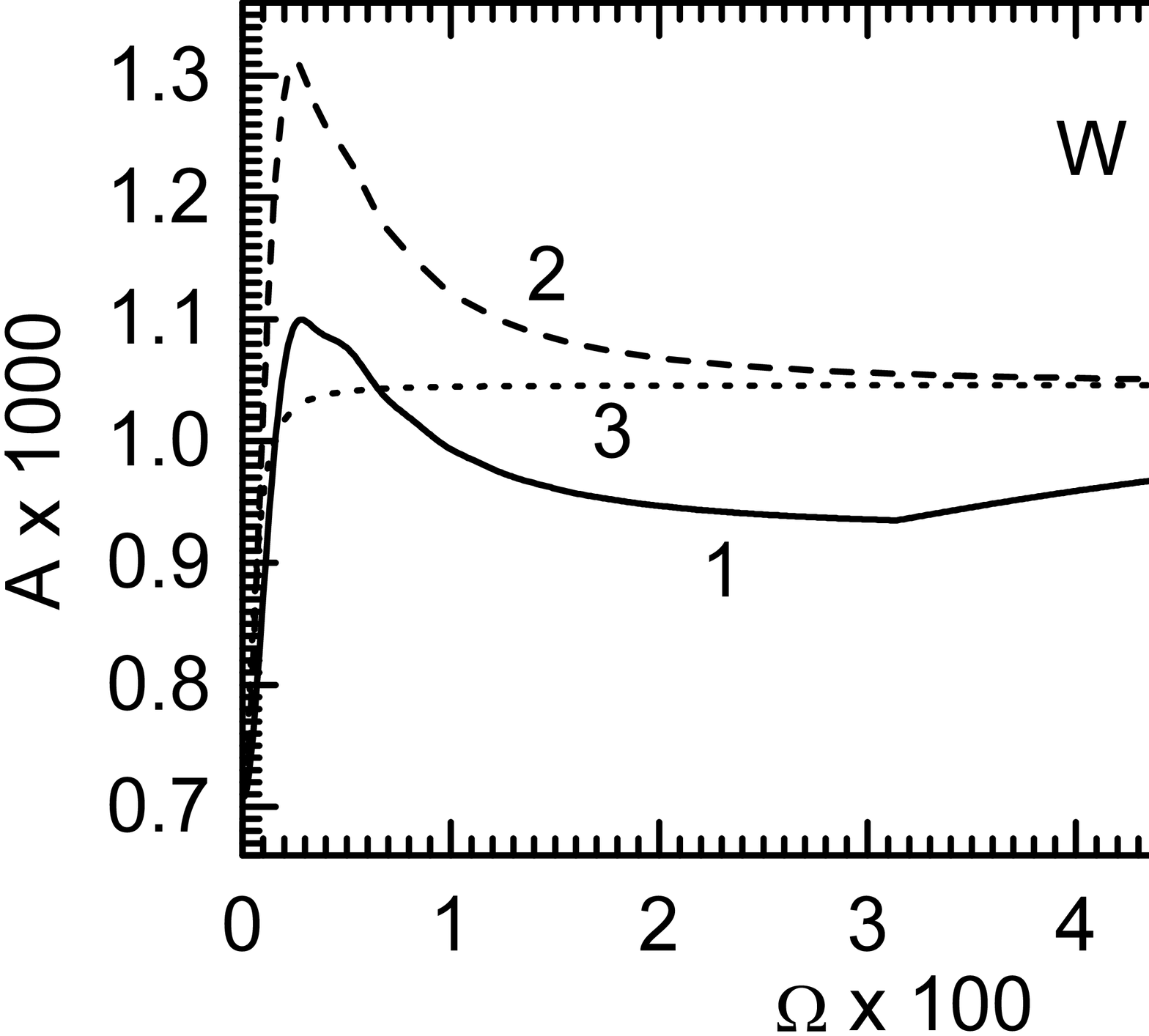,width=.5\textwidth,angle=0}

\vspace*{-27mm}

\caption{The reflectance $R$ (left plots) and absorptance $A$ (right
plots) as functions of $\Omega$ at $\beta=2.83\cdot{}10^{-3}$,
$\gamma=10^{-3}$, $r=1.07$, $\theta=60^{\circ}$, $\varepsilon_1=1$,
$\varepsilon_2=2$, $W=200$ (upper plots), $W=1000$ (lower plots):
\ \ 1 -- quantum plasma (solid line), \ 2 -- classical spatial
dispersion case (dashed line), \ 3 -- Drude -- Lorentz approach (dotted
line).}
\label{fig:raodcq}

\end{figure}

The numerical studies have shown that for the frequencies $\Omega\gg%
\pi/W$ (or $\omega\gg\pi{}v_F/d$), the power coefficients $R$, $T$ and
$A$ for quantum plasma {\bf almost coincide} with those in cases of
both the classical spatial dispersion plasma and the Drude -- Lorentz
approach. So the results for the quantum plasma presented at the
figs.~\ref{fig:raq_s}~--~\ref{fig:cfthq}, are valid also in cases of
the classical and the Drude -- Lorentz approaches for the frequencies
$\Omega\gtrsim{}0.1$. However for the frequencies $\Omega\sim\pi/W$
($\omega\sim\pi{}v_F/d$) when the film width satisfies the condition
$W\gtrsim\beta^{-1}$ ($d\gtrsim{}c/\omega_p$), one observes a difference
of the power coefficients evaluated for various approaches. Typical
results illustrating such a disagreement are presented at the fig.%
~\ref{fig:raodcq} for the values $W=200$ and $1000$ ($d=25.72$~nm and
$128.6$~nm). One sees that the deviation is the most visible for the
absorptance $A$ and is more clear for large values $W$. Hence for these
frequencies, both the quantum wave and the spatial dispersion effects of
degenerate electron plasma contribute to interaction of the plasma with
electromagnetic wave.

\begin{figure}[ht]

\vspace*{27mm}
\hspace*{-1mm}
\epsfig{file=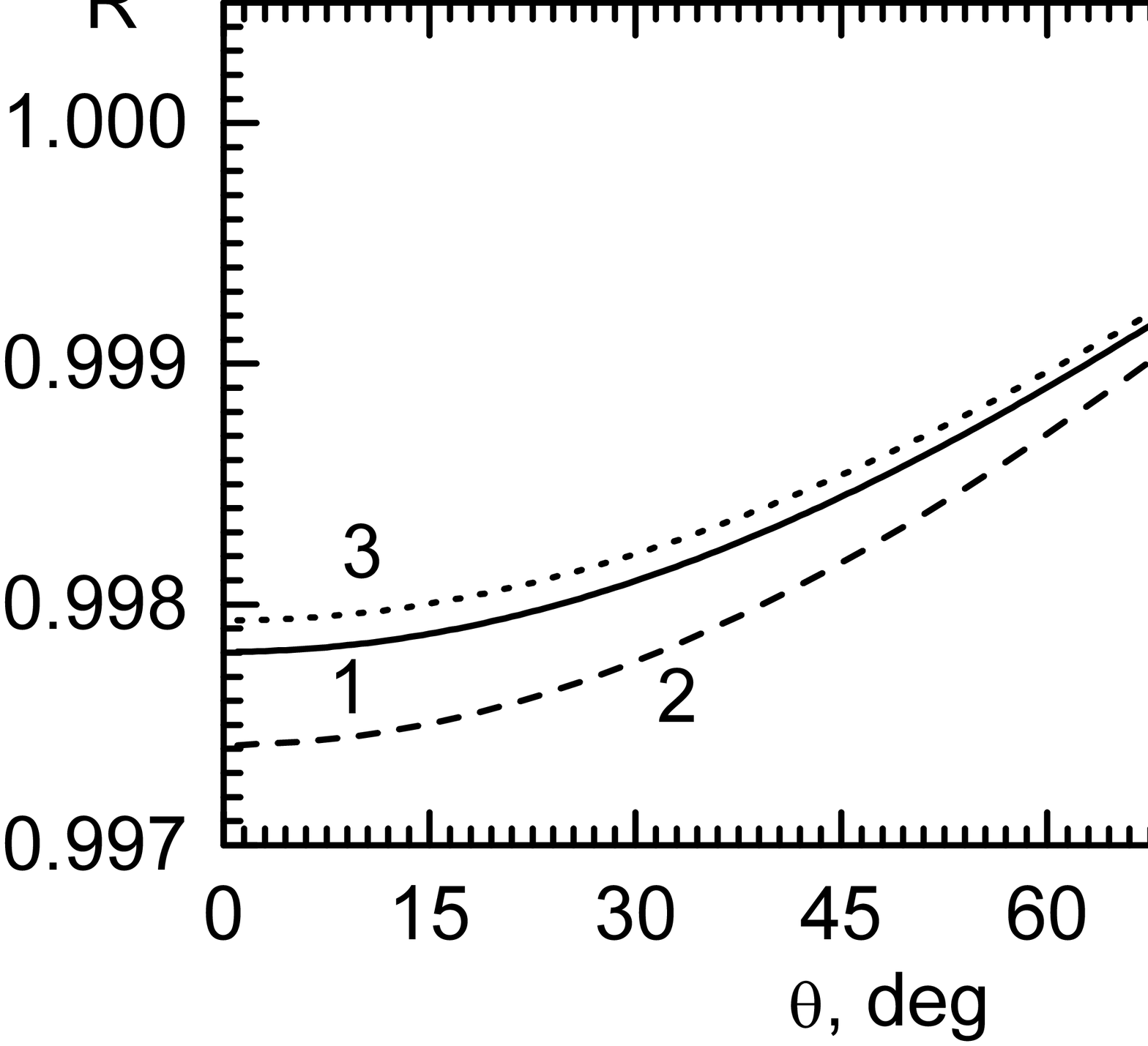,width=.5\textwidth,angle=0}
\hspace*{-2mm}
\epsfig{file=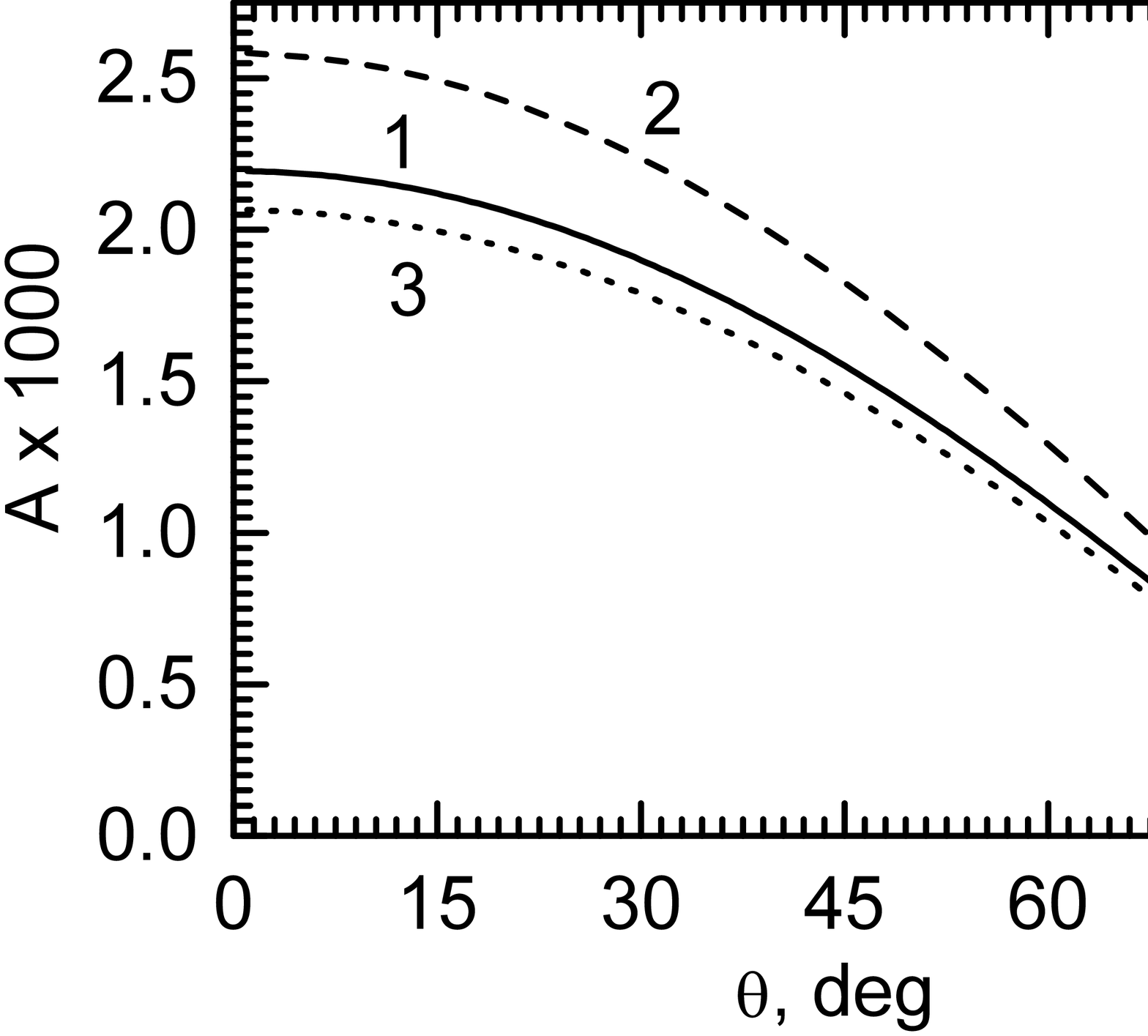,width=.5\textwidth,angle=0}

\vspace*{-27mm}

\caption{The reflectance $R$ (left plot) and absorptance $A$ (right
plot) as functions of $\theta$ at $\beta=2.83\cdot{}10^{-3}$,
$\gamma=10^{-3}$, $r=1.07$, $\Omega=\pi/W$, $\varepsilon_1=1$,
$\varepsilon_2=2$, $W=1000$: \ \ 1 -- quantum plasma (solid line),
\ 2 -- classical spatial dispersion case (dashed line), \ 3 -- Drude
-- Lorentz approach (dotted line).}
\label{fig:rathdcq}

\end{figure}

The disagreement of various approaches is shown also at the fig.%
~\ref{fig:rathdcq} for reflectance $R$ and absorptance $A$ as functions
of incidence angle $\theta$, evaluated at the frequency $\Omega~=~\pi/W$.
One sees the difference at small incidence angle. And the results for
the classical spatial dispersion plasma deviate from the Drude --
Lorentz approach stronger than for the quantum plasma. But when the
$\theta$ increases to $90^{\circ}$, the disagreement decreases and
vanishes.

These results agree with the data presented in the papers
\cite{KlFu3,PDMP} in cases of the classical spatial dispersion and the
Drude -- Lorentz approaches. The disagreement of various approaches
takes place for the frequencies $\omega$ covering the terahertz and
infrared ranges. As it was shown in \cite{KlFu3}, the difference of the
spatial dispersion dielectric functions from the Drude -- Lorentz one
for the frequencies $\omega\sim\pi{}v_F/d$ is a manifestation of the
resonance at the frequency of the periodic motion of electrons across
the film between its borders which is just equal to $\pi{}v_F/d$. And
also, a difference of the approaches is the most visible just for the
absorptance $A$ when the film width $d\gtrsim{}c/\omega_p$ since the
electromagnetic wave is well damped beyond the skin depth $c/\omega_p$.

The discrepancy of the power coefficients can be explained also by a
contribution of the values $Q_n$ by (\ref{qqx}) to the surface impedance
(\ref{sf_imp2}) in cases of the quantum and classical spatial dispersion
dielectric functions (\ref{epqu_tr}) and (\ref{epcl_tr}), for the values
$n\sim{}1$ (see \cite{PDMP}). First of all, for the values $W\gtrsim%
\beta^{-1}$, a contribution of first terms with $n\ne{}0$ to
(\ref{sf_imp2}) is essential. Further for the values $\Omega\sim\pi{}%
n/W\sim{}Q_n$ in case of small $n$ numbers, one has a deviation of both
the quantum and the classical spatial dispersion transverse dielectric
functions from the Drude -- Lorentz one owing to smoothed critical
behavior of functions (\ref{b12}).

\begin{figure}[ht]

\vspace*{27mm}
\hspace*{0mm}
\epsfig{file=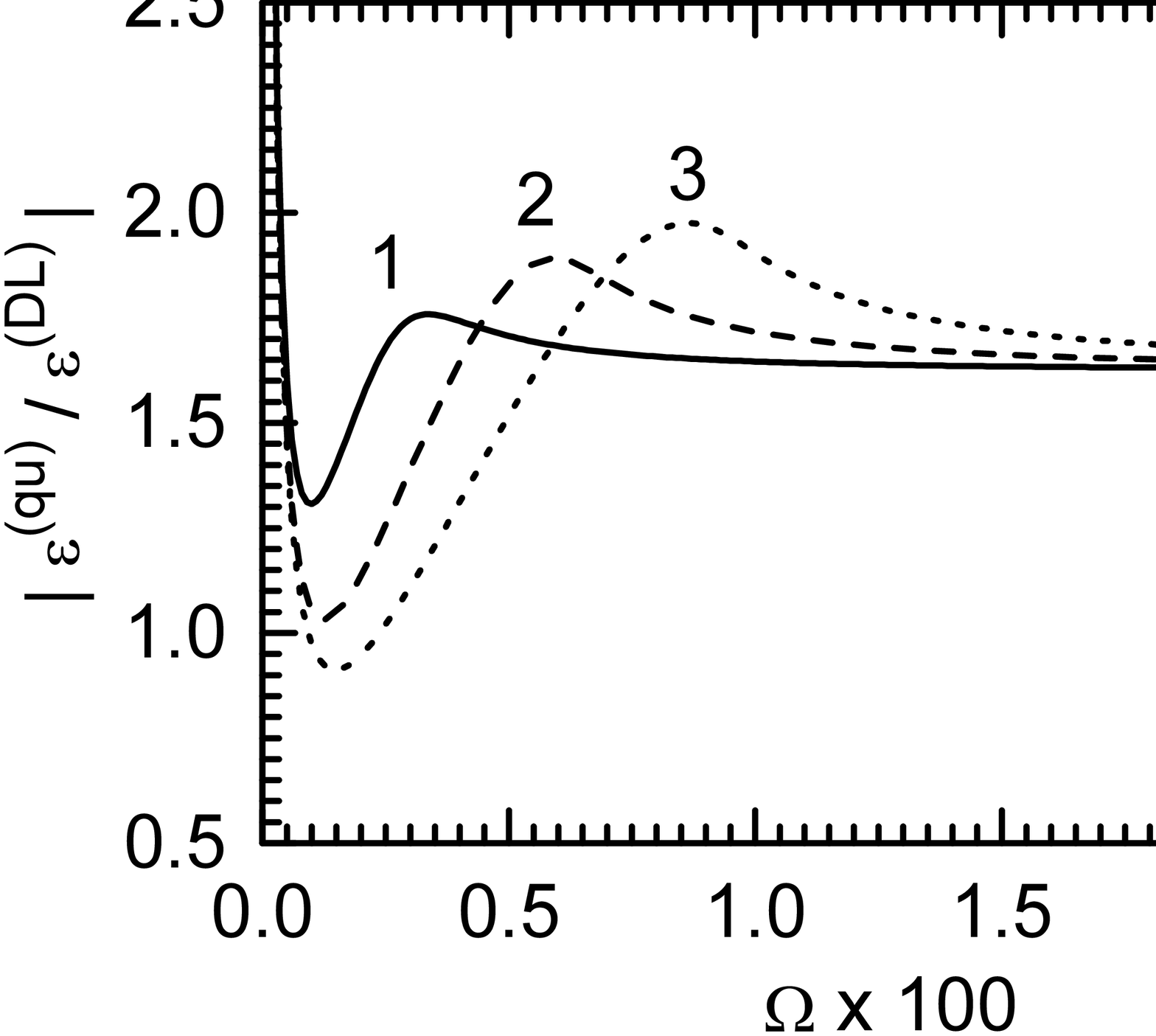,width=.5\textwidth,angle=0}
\hspace*{-3mm}
\epsfig{file=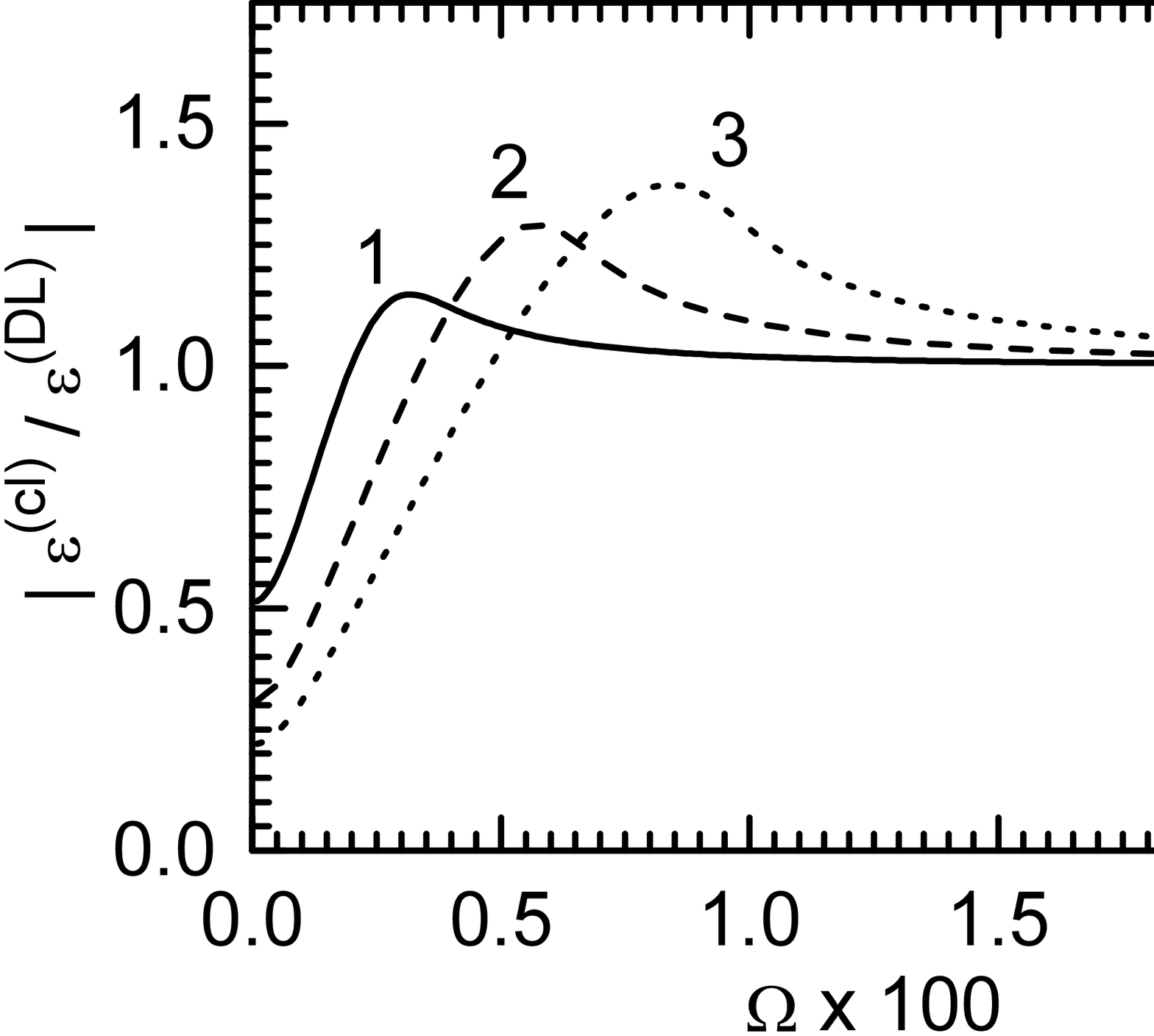,width=.5\textwidth,angle=0}

\vspace*{-25mm}

\caption{The ratios $|\varepsilon^{(qu)}_{tr}/\varepsilon^{(DL)}_{tr}|$
(left plot) and $|\varepsilon^{(cl)}_{tr}/\varepsilon^{(DL)}_{tr}|$
(right plot) as functions of $\Omega$ with $Q=Q_n$ at $\beta=2.83%
\cdot{}10^{-3}$, $\gamma=10^{-3}$, $r=1.07$, $\theta=60^{\circ}$,
$\varepsilon_1=1$, $W=1000$: \ \ 1~--~$n=1$ (solid line), \ 2~--~$n=2$
(dashed line), \ 3~--~$n=3$ (dotted line).}
\label{fig:mepdqc}

\end{figure}

At the fig.~\ref{fig:mepdqc}, such a deviation is demonstrated for the
magnitudes of the dielectric functions. We present the values
$|\varepsilon^{(qu)}_{tr}/\varepsilon^{(DL)}_{tr}|$ and
$|\varepsilon^{(cl)}_{tr}/\varepsilon^{(DL)}_{tr}|$ as functions of
$\Omega$ with $Q=Q_n$ by (\ref{qqx}) where $Q_x$ is evaluated according
to (\ref{qx2}), for first three values $n$. One sees that the extremes
(maxima) of the ratios take place in the points $\Omega\simeq\pi{}n/W$.
And in the vicinity of these extremal values, the ratios differ from
the $1$. It can be seen also that the quantum dielectric function
differs from the classical spatial dispersion one in the considered
frequency region. Such a difference is caused by a contribution of the
values $rQ^2_n/2$ and therefore, is explained by the influence of the
quantum wave properties of the electrons on the dielectric function.

\section*{Conclusion}

In the paper, we have investigated numerically an interaction of the
electromagnetic S-wave with thin flat metallic film placed between two
dielectric media, in the framework of the quantum degenerate electron
plasma approach. We have selected for investigation the reflectance,
transmittance and absorptance power coefficients. We have taken the
transverse dielectric function of the quantum degenerate electron plasma
with constant relaxation time in the Mermin approach in case of specular
reflection of electrons from the film borders. The power coefficients
were studied as functions of the frequency or the incidence angle
at various nanoscale widths of the film and various second transparent
media or substrate materials. The obtained results for the power
coefficients of the quantum plasma were compared with the coefficients
evaluated both in the classical spatial dispersion plasma and in the
Drude -- Lorentz approach without spatial dispersion.

It was shown that in the wide frequency interval from the terahertz
range to ultraviolet one, the power coefficients of the quantum plasma
have a smooth behavior without critical peaks. Further, dependence of
the coefficients on the incidence angle in case of S-wave is reminiscent
to the dependence in the P-wave case. And also for the large enough
frequencies, the S-wave power coefficients of the quantum plasma almost
coincide with the coefficients in cases of both the classical spatial
dispersion plasma and the Drude -- Lorentz approach. But for small
frequencies lying in the terahertz and infrared ranges, the power
coefficients evaluated in cases of various approaches differ from
each other at the nanoscale width of the film. The origin of such a
disagreement of the power coefficients was discussed. A discrepancy
of the power coefficients indicates a manifestation of both the
quantum wave property of electrons in plasma and the dimensional
effects of interaction the electromagnetic wave with plasma due to
the spatial dispersion.

The obtained results should be used in the theoretical studies of
the quantum plasma as well as the interaction of the electromagnetic
wave with nanoscale conducting objects. The results may have also a
practical application in case of creation and use of optical
instruments having thin metallic material.

\bigskip

This work is supported by the Research Grant of the President of Russian
Federation \ {\it MK-7359-2016-9} \ and by the RFBR Grants \
{\it 14-07-90009 Bel\_a}, \ {\it 14-47-03608 r\_centr\_a}.

\end{document}